**Magnetic multipoles and correlation shortage in pyrochlore cerium stannate** ($Ce_2Sn_2O_7$)

S. W. Lovesey and G. van der Laan

*Diamond Light Source, Didcot, Oxfordshire OX11 0DE, United Kingdom*

**Abstract**. Cerium electronic multipoles in trigonal symmetry viewed in magnetic neutron diffraction are investigated. Analytic expressions for all the magnetic multipoles, and radial integrals calculated with an established atomic code, are used to scrutinize a recent claim to have observed an octupole-ice configuration in a powder sample of pyrochlore $Ce_2Sn_2O_7$. Three equally plausible suites of multipoles belonging to uncorrelated cerium ions give equally satisfactory accounts of the available diffraction pattern. Our multipoles are suitable for future experiments using Bragg diffraction from a compound that supports long-range magnetic order.

## I. INTRODUCTION

Studies of the magnetic properties of rare-earth materials surged in the 1950s. The suggestion that rare-earth Kramers ions in cubic symmetry could provide suitable maser materials was one motivating factor [1]. The celebrated Judd-Ofelt theory of optical absorption appeared shortly thereafter [2, 3]. Today, rare earth elements are crucial components of many technologies, including, batteries, catalysts, permanent magnets, and electronic devices. A significant feature of early experimental studies of rare earth materials was the exploitation of magnetic neutron diffraction. Indeed, numerous successful studies by Koehler and collaborators, at Oak Ridge National Laboratory, promoted the technique to its present status of the method of choice for the determination of magnetic motifs [4]. Notably, the study by Koeller and Wollan [5] of paramagnetic compounds containing Er and Nd inspired Trammell in 1953 [6] to produce a complete theory of the neutron magnetic scattering amplitude (George Trammell was Richard Feynman's first postgraduate student). His pioneer theory was vastly improved in 1966 when Johnston proved that the spin and orbital components of the amplitude contain the same radial integrals, whereas Trammell used seemingly quite different radial integrals for each of the two components [7, 8]. Alongside these theoretical developments were numerous studies of resonance phenomena (EPR, NMR) and the Mössbauer effect, making extensive use of Racah algebra and the concept of operator equivalents introduced by Elliott and Stevens in the year Trammell published his account of magnetic neutron scattering [9, 10, 11, 12]. Likewise, there were big strides in our knowledge of actinide materials [13].

Many probes of magnetic materials, including, resonance techniques, magnetic circular dichroism, and x-ray and neutron scattering, yield information on electronic multipoles [14, 15, 16, 17]. Familiar dipole operators are the magnetic moment **μ** = (**L** + 2**S**) and the spin anapole (**S** × **R**), where **L**, **S** and **R** are operators for orbital angular momentum, spin and position.

An aim of the present study is to scrutinize a recent claim that octupoles are the predominant magnetic entity of cerium in pyrochlore $Ce_2Sn_2O_7$ at a low temperature (T = 0.05 K) [18]. The claim, which amounts to an observation of octupole-ice, is predicated on an analysis of extensive diffuse magnetic neutron diffraction gathered on a powder sample. Our method of analysis is entirely different. Specifically, we have calculated cerium multipoles allowed in trigonal symmetry $\bar{3}m$ and $\bar{3}m'$. To this end, we use a $Ce^{3+}$ wavefunction inferred from neutron spectroscopy, and implement the precise definition of octupole-ice by selecting appropriate parameters [18]. Neutron scattering by tripositive cerium in a cubic environment has been thoroughly investigated in a recent theoretical paper by Ayuel *et al*. [19].

Multipoles in the neutron scattering amplitude depend on the magnitude of the scattering wavevector, which appears in the argument of spherical Bessel functions averaged over the electronic radial density. Authors of reference [18] follow in the steps of Trammell [6] and are content with the 4f radial density of a hydrogen ion, which renders radial integrals in the scattering amplitude a function of an effective core charge, $Z_c$. While $Z_c$ can be correlated with independent observables, e.g., the spin-orbit coupling constant [20] and crystal-field parameters [21], it is tantamount to a free parameter. Trammell uses $Z_c$ = 23 ($Er^{3+}$), $Z_c$ = 20 ($Nd^{3+}$), while $Z_c$ = 17 ($Ce^{3+}$) yields a favourable fit to the available diffraction pattern [18]. By contrast, we adopt a modern approach and calculate the electronic density for a free ion using a tried and tested atomic code [22, 23, 24], with no attempt to simulate solid-state effects that are likely very small since 4f electrons form an "inner shell" shielded within 5d and 6s orbitals. Significant differences between the radial density of the 4f hydrogen ion and those calculated with atomic codes are now well established [25, 26], and we corroborate the findings.

Diffraction is attributed to uncorrelated cerium ions as in a simple paramagnetic material. Satisfactory agreement is found between the minimal model and existing experimental data on $Ce_2Sn_2O_7$, leaving the apparent shortage of magnetic correlations unanswered. A radial density of the 4f hydrogen ion with $Z_c$ = 17 [18] gives a marginal improvement between the model and data. Specifically, intensity calculated for large wavevectors, for which there are few data, is weaker with a hydrogen radial density that is more compact than the free ion calculated with an atomic code and relativistic corrections. Data on offer do not rule on the existence of octupole-ice. Our cerium multipoles, the essence of a pyrochlore scenario, show that the definition of octupole-ice [18] amounts to minimizing diagonal components of multipoles and, specifically, the dipole is very small. However, off-diagonal components of higher-order multipoles, allowed by the triad axis of symmetry in cerium sites, maintain their significant values. In consequence, diffraction patterns are essentially unchanged upon implementing the definition of octupole-ice. Site symmetry allows six higher-order multipoles and the magnitude of the triakontadipole rivals that of the octupole. Many of the multipoles, including the triakontadipole, appear to be missing in the analysis reported in reference [18]. Our findings make a case for additional measurements at large scattering wavevectors.

## II. MATERIAL PROPERTIES

Phase-pure $Ce_2Sn_2O_7$ is paramagnetic. The measured susceptibility is consistent with a tripositive cerium free ion and a magnetic moment $g\sqrt{[J(J+1)]} = 2.54$ at high temperatures ($Ce^{3+}$, $4f^1$, $^2F$, $J = 5/2$ and Landé factor $g = 6/7$) [27]. Susceptibility measurements at moderate temperatures, 1 K < T < 10 K, imply an effective magnetic moment $= \sqrt{\langle\mu^2\rangle} \approx gM = 9/7$, using a magnetic projection $M = 3/2$, while there is no sign of long-range magnetic order down to 0.02 K [28]. This contrasts with $Nd_2Zr_2O_7$ that also hosts a Kramers ion ($Nd^{3+}$, $^4I$, $J = 9/2$), with long-range antiferromagnetic order below $\approx 0.4$ K and evidence of dipole-octupole character [29].

The chemical structure of $Ce_2Sn_2O_7$ is illustrated in Fig. 1. Ce ions occupy centrosymmetric sites (16d with site symmetry $D_{3d}$; 1/2, 1/2, 1/2) in space group $Fd\bar{3}m$ (#227) [27]. A Kramers doublet $|u\rangle$ and $|\bar{u}\rangle$ of $Ce^{3+}$ is built from atomic states $|J, M\rangle$ with $J = 5/2$, $J' = 7/2$ and projection $M = 3/2$. The conjugate function $|\bar{u}\rangle$ is derived from $|u\rangle$ by the rule $|k, \mu\rangle \to (-1)^{k-\mu}|k, -\mu\rangle$ that is our definition of a time-reversed state [12, 16, 17]. Analyse of the energy spectrum collected by neutron spectroscopy yields a normalized wavefunction,

$$|u\rangle = a|J, M\rangle + b|J, -M\rangle + \alpha|J', M\rangle, \qquad (1)$$

with $a \approx 0.87$, $b \approx 0.46$ and $\alpha \approx -0.15$ [18]. Octupole-ice is defined by $a = b$ [18]. An effective magnetic moment $= \sqrt{\langle\mu^2\rangle} \approx gM = 9/7$.

The ground state,

$$|g\rangle = [|u\rangle + i|\bar{u}\rangle]/\sqrt{2}, \qquad (2)$$

obeys site symmetry $D_{3d}$ for real coefficients in $|u\rangle$. We set $\alpha = 0$ for this special case, and choose $(a^2 + b^2) = 1$ with $\langle g|g\rangle = 1$. All other calculations, based on $|u\rangle$ alone, include an admixture of the two manifolds with $\alpha \approx -0.15$.

## III. MULTIPOLES

Axial magnetic multipoles are labelled $\langle t^K_Q \rangle$ or $\langle T^K_Q \rangle$ for one of two point-groups, while Dirac multipoles that are both time-odd and parity-odd are forbidden. (In a discussion of neutron scattering by a pyrochlore it is worth noting that a Dirac monopole (K = 0) is forbidden [17, 30]). In our notation, $\langle ... \rangle$ denotes the ground-state, or time-average, value of the enclosed spherical tensor operator, integer K is the rank, and (2K + 1) projections obey $-K \leq Q \leq K$.

Consider axial multipoles $\langle t^K_Q \rangle$ prescribed by symmetry $D_{3d}$ ($\bar{3}m$). Principal axes ($\xi$, $\eta$, $\zeta$) are taken to be $\xi = [-1, -1, 2]/\sqrt{6}$, $\eta = [1, -1, 0]/\sqrt{2}$, and $\zeta = [1, 1, 1]/\sqrt{3}$. The triad axis of symmetry parallel to $\zeta$-axis requires $Q = \pm 3n$, with n a positive integer. In addition, the diad axis of symmetry in $D_{3d}$ requires $2_\eta \langle t^K_Q \rangle = (-1)^{K+Q} \langle t^K_{-Q} \rangle = \langle t^K_Q \rangle$, where $2_\eta$ represents two-fold rotation about the η-axis. In consequence, $\langle t^K{}_0 \rangle = 0$ for K odd. Using the Hermitian property of our multipoles yields the general result $(-1)^K \langle t^K{}_Q \rangle^* = \langle t^K{}_Q \rangle$, and $\langle t^K{}_Q \rangle$ is purely imaginary for K odd. Radial integrals mentioned in §I are functions of the wavevector κ, and they are denoted by $\langle j_m(\kappa) \rangle$ with m = 2, 4 and 6. Upon using Eq. (2), the identity $\langle u|\mathbf{t}^K|u \rangle = - \langle \bar{u}|\mathbf{t}^K|\bar{u} \rangle$, and reduced matrix-elements listed in the Appendix,

$$\langle t^3{}_{+3} \rangle = \langle g|t^3{}_{+3}|g \rangle = i(4/7) \sqrt{(1/35)}\, h(\kappa),$$

$$\langle t^5{}_{+3} \rangle = -i(5/11) \sqrt{(1/231)} \{\langle j_4(\kappa) \rangle + 12 \langle j_6(\kappa) \rangle\}, \qquad (3)$$

where $h(\kappa) = \{\langle j_2(\kappa) \rangle + (10/3) \langle j_4(\kappa) \rangle\}$. Results for $\langle t^K{}_Q \rangle$ are independent of the specific values of coefficients a and b in $|u \rangle$, i.e., multipoles in question are symmetry-protected. Even rank multipoles are absent when the admixture of J′ = 7/2 is neglected (α = 0). (The multipoles represent entanglement of anapole and spatial degrees of freedom [31].) Multipoles referred to crystal axes $\langle \tau^K{}_Q \rangle$ are purely real, e.g., $\langle \tau^3{}_{+2} \rangle = -i\langle t^3{}_{+3} \rangle/\sqrt{3}$, $\langle \tau^5{}_{+2} \rangle = +i\langle t^5{}_{+3} \rangle/\sqrt{3}$, $\langle \tau^K{}_0 \rangle = 0$ for K odd, together with $\langle \tau^5{}_{\pm 4} \rangle = 0$.

Site symmetry $D_{3d}$ must be amended to yield a magnetic dipole. Point-group $\bar{3}m'$ qualifies as a candidate, because $2_{\eta}' \langle T^K{}_Q \rangle = (-1)^{K+1} \langle T^K{}_Q \rangle^* = \langle T^K{}_Q \rangle$. In this case, odd (even) rank multipoles are purely real (imaginary). By way of an example, sites 16d in space-group $Fd\bar{3}m'$ (#227.131) possess symmetry $\bar{3}m'$. The cubic structure does not possess long-range magnetic order, and allowed Bragg spots have not been observed. Calculations using Eq. (1) show that the wavefunction is consistent with $\bar{3}m'$, namely,

$$\langle T^1{}_0 \rangle = \langle u|T^1{}_0|u \rangle = (a^2 - b^2) (3/7) \{\langle j_0(\kappa) \rangle + (8/5) \langle j_2(\kappa) \rangle\}$$

$$+ a\alpha\, (2/21) \sqrt{(10)} \{\langle j_0(\kappa) \rangle - (1/2) \langle j_2(\kappa) \rangle\} + \alpha^2 (4/7) \{\langle j_0(\kappa) \rangle + (2/3) \langle j_2(\kappa) \rangle\},$$

$$\langle T^3{}_0 \rangle = - (a^2 - b^2) (2/5) \sqrt{(1/7)}\, h(\kappa) - \alpha^2 (4/3) \sqrt{(1/7)} \{\langle j_2(\kappa) \rangle + (4/11) \langle j_4(\kappa) \rangle\},$$

$$\langle T^3{}_{+3} \rangle = - ab\, (8/7) \sqrt{(1/35)}\, h(\kappa) + \alpha b\, (2/21) \sqrt{(2/7)} \{\langle j_2(\kappa) \rangle - (3/4) \langle j_4(\kappa) \rangle\},$$

$$\langle T^4{}_{+3} \rangle = i\alpha b\, (7/22) \sqrt{(21/10)} \langle j_4(\kappa) \rangle,$$

$$\langle T^5{}_0 \rangle \approx - (a^2 - b^2) (5/77) \sqrt{(5/33)} \{\langle j_4(\kappa) \rangle + 12 \langle j_6(\kappa) \rangle\}$$

$$- a\alpha\, (6/77) \sqrt{(6/11)} \{\langle j_4(\kappa) \rangle - (5/6) \langle j_6(\kappa) \rangle\},$$

$$\langle T^5{}_{+3} \rangle \approx ab\, (10/11) \sqrt{(1/231)} \{\langle j_4(\kappa) \rangle + 12 \langle j_6(\kappa) \rangle\}$$

$$- \alpha b\, (1/11) \sqrt{(6/385)} \{\langle j_4(\kappa) \rangle - (5/6) \langle j_6(\kappa) \rangle\},$$

$$\langle T^6_{+3} \rangle = -i\alpha b\,(35/22)\,\sqrt{(1/3)}\,\langle j_6(\kappa) \rangle. \tag{4}$$

It is notable that $\langle T^3_0 \rangle$ does not contain an interference term $\propto (a\alpha)$, which is found to be identically zero (the Clebsch-Gordan coefficient $(J\ -M, J'\ M|30) = 0$), whereas $\langle T^5_0 \rangle$ possesses such a term. Terms proportional to $\alpha^2$ are neglected in results for triakontadipoles, because they are very small.

## IV. RADIAL INTEGRALS
### A. Hydrogen ion

Early studies of rare earth ions used a primitive radial wavefunction $\propto \{R^3 \exp(-Z_c R/2a_o)\}$ [6], where $a_o$ is the Bohr radius. An interesting correlation between neutron diffraction data of interest and studies of the spin-orbit coupling constant is achieved with the relation $Z_c^3 \approx \langle (13.90\,a_o/R)^3 \rangle$. Using $\langle (a_o/R)^3 \rangle = 4.462$ for $Ce^{3+}$ ($Z = 58$) yields an effective core charge $Z_c \approx 22.9$ [26, 32]. Judd and Lindgren [20] cite an empirical formula $Z_c = (Z - \mathcal{S})$ with $\mathcal{S} = 34$ or 35 which they derive from an analysis of spin-orbit constants. In addition, the authors investigate the merits of a modified hydrogen wavefunction to provide atomic parameters. Ayuel *et al*. [19] report $\langle (R/a_o)^3 \rangle = 1.835$ from an Hartree-Fock calculation for the $Ce^{3+}$ wavefunction [25], and this value is reproduced by a hydrogen ion with $Z_c \approx 16.3$. The numerical examples of the effective core charge given here, and additional values in foregoing text, demonstrate that $Z_c$ has no specific physical significance, and, as such, a free parameter.

Radial integrals are functions of $\kappa = [4\pi \sin(\theta)/\lambda]$. A subscript (o) on radial integrals calculated with a hydrogen wavefunction distinguishes them from more reliable estimates that are the subject of the following sub-section. Results for radial integrals displayed in equation (37) in reference [6] contain a few typographical errors. We find,

$$\langle j_0(\kappa) \rangle_o = [1 + q^2]^{-8}\,[1 - 7q^2 + 7q^4 - q^6],$$

$$\langle j_2(\kappa) \rangle_o = 6q^2\,[1 + q^2]^{-8}\,[1 - (10/7)\,q^2 + (5/21)\,q^4],$$

$$h_o(\kappa) = (2/3)\,q^2\,[1 + q^2]^{-8}\,[9 + 50\,q^2 - 15\,q^4], \tag{5}$$

with $q = (2/3)(w/Z_c)$, and the dimensionless variable $w = 3a_o\kappa$. The first zero of $\langle j_0(\kappa) \rangle_o$ occurs at $q \approx 0.4142$. The radial integral $\langle j_2(\kappa) \rangle_o$ has its principal maximum at $q \approx 0.3367$, with $\langle j_2(0.337) \rangle_o \approx 0.242$, while the factor $h_o(\kappa) = \{\langle j_2(\kappa) \rangle_o + (10/3)\langle j_4(\kappa) \rangle_o\}$ in the octupoles is a maximum for $q \approx 0.4812$ at which it achieves the value $\approx 0.577$.

### B. Atomic calculations

Radial integrals $\langle j_m(\kappa) \rangle$ for $Ce^{3+}$ in our study are calculated using a $[Xe]4f^1$ wavefunction obtained from Cowan's atomic Hartree-Fock program with relativistic corrections [22, 23]. This code has proven to give an excellent agreement with x-ray spectroscopic data in the case of all rare-earth elements for all ionization states known in the solid state [24]. Our radial integrals, some of which are displayed in Fig. 2, agree nicely with the limited set of values published by Freeman and Desclaux [26]. Applied to $Ce^{2+}$ our program returns radial integrals in agreement with a standard tabulation [33].

Using $Z_c = 17$ [18] the maximum in $h_o(\kappa)$ occurs at $\kappa \approx 7.73$ Å$^{-1}$, while $h(\kappa)$ we calculate has its maximum at $\kappa \approx 7.95$ Å$^{-1}$. The similarity between $h_o(\kappa)$ for $Z_c = 17$ and our $h(\kappa)$, illustrated in Fig. 2, is reason to choose the quoted value of $Z_c$, for it ensures a tolerable representation of $\langle T^3{}_Q \rangle$ using a hydrogen-like radial density. However, the two calculations, $h_o(\kappa)$ and $h(\kappa)$, are significantly different despite almost identical positions for their principal maxima, because the hydrogen-like radial density is too compact. Values of $\langle j_0(\kappa) \rangle_o$ and $\langle j_2(\kappa) \rangle_o$ with $Z_c \sim 25.3$ and $Z_c \sim 17.2$, respectively, reproduce principal features of our corresponding radial integrals. Fig. 2 displays radial integrals m = 0, 2 and 4 derived from a 4f hydrogen wavefunction using $Z_c = 17$ [18] and $Z_c = 25.3$, together with our corresponding radial integrals. The dipole $\langle T^1{}_0 \rangle$, being dominated by the radial integral m = 0 for dissimilar coefficients a and b, is poorly represented by $\langle j_0(\kappa) \rangle_o$ and $Z_c = 17$. A superior representation requires a much larger effective core charge, as seen with $\langle j_0(\kappa) \rangle_o$ and $Z_c = 25.3$ in Fig. 2. Hereafter, we report intensities calculated with our radial integrals.

## V. DIFFRACTION PATTERN

An average of the neutron cross-section with respect to directions of the scattering wavevector yields a diffraction pattern,

$$I = (1/4\pi) \int d\hat{\boldsymbol{\kappa}} \{\langle \mathbf{Q}_\perp \rangle \cdot \langle \mathbf{Q}_\perp \rangle\} = \sum_{K,Q} [3/(2K + 1)] |\langle T^K{}_Q \rangle|^2 + \sum_{K',Q'} [3/(K' + 1)] |\langle T^{K'}{}_{Q'} \rangle|^2, \quad (6)$$

with K even and K' odd. Allowed multipoles are K even (= 2, ..., 2l) and K' odd (= 1, 3, ..., 2l + 1) [17]. However, K' = 7 does not occur for 4f$^1$. Eq. (6) is the analogue of Trammell's result in Eq. (8) [6].

Figure 3 shows the diffraction pattern calculated from Eq. (6) using the two multipoles Eq. (3) for $D_{3d}$ calculated with our radial integrals. Calculated diffraction is a satisfactory description of the experimental data [18] that extend to $\kappa \approx 9$ Å$^{-1}$ [18]. Beyond is the region where hydrogen-like ($Z_c = 17$) and calculated radial densities are distinctly different, as we see in Fig. 2. These neutron diffraction patterns are difference data that do not contain the dipole contribution allowed in $\bar{3}$m′. Specifically, data displayed in Figs. 3-5 (blue solid squares) are the difference between patterns measured at 5 K and 0.05 K [18]. Figure 4 shows our multipoles Eq. (4) evaluated with a = 0.87, b = 0.46 and α = − 0.15 and the corresponding diffraction pattern. Intensity labelled non-dipole (red curve) is Eq. (6) omitting the dipole that is displayed separately as a black curve, i.e., the red curve simulates the difference data. The choice a = b for parameters in Eq. (1) defines the octupole-ice structure [18]. For this choice of parameters, diagonal multipoles $\langle T^K{}_0 \rangle$ are small compared to off-diagonal multipoles $\langle T^K{}_{+3} \rangle$, as we see in Fig. 5. The small dipole contribution shown separately by a black curve lifts our calculation to a better fit to data. By contrast, diagonal and off-diagonal multipoles for a = 0.87, b = 0.46 and α = − 0.15 in Fig. 4 have comparable magnitudes. And, $2\langle T^5{}_{+3} \rangle \sim - \langle T^3{}_{+3} \rangle$ in all cases, so the triakontadipole is not to be set aside. Even-rank multipoles, created by J-mixing in Eq. (1) [31] are comparatively small and are much the same in Figs. 4 and 5. Diffraction patterns displayed in Figs. 3, 4 and 5 possess a maximum at $\kappa \approx 8$ Å$^{-1}$, which is a good match to the experimental data.

## VI. DISCUSSION AND CONCLUSIONS

In summary, three physically justified suites of multipoles for uncorrelated cerium ions yield equally useful descriptions of recently published diffraction patterns for pyrochlore cerium stannate [18]. The suites belong to two classes defined by candidate symmetries of cerium sites, $D_{3d}$ ($\bar{3}$m) and $\bar{3}$m′, and one of the three follows the definition of octupole-ice devoid of magnetic correlations.

Satisfactory agreement between experimental diffraction data for pyrochlore $Ce_2Sn_2O_7$ and a theory reported by Sibille *et al*. [18] owes much to their elected effective core charge $Z_c$ in a hydrogen-like radial density of $Ce^{3+}$. There is no independent evidence to corroborate $Z_c$ = 17 which the authors use. E.g., an empirical formula $Z_c = (Z - \mathcal{S})$ with $\mathcal{S}$ = 34 or 35, derived from an analysis of spin-orbit constants [20], suggests a larger $Z_c$ in line with earlier diffraction studies with $Z_c$ = 23 ($Er^{3+}$) and $Z_c$ = 20 ($Nd^{3+}$) [6]. Our calculations for the free ion $Ce^{3+}$ use Cowan′s code with relativistic corrections [22, 23, 24]. Radial integrals accord with independent calculations [26], and we confirm that the hydrogen-like radial density is overly compact.

In consequence, we have confidence in our prediction of diffraction at scattering wavevectors beyond the modest range covered in the available experiments. It is a range accessed with current diffraction instruments, including POLARIS at the ISIS Facility, UK. Convolution of our prediction with magnetic correlations might alter the story that has emerged, but useful descriptions of measured diffraction patterns in the modest range of wavevectors reported in Figs. 3, 4 and 5 suggest otherwise.

There are six multipoles in addition to the dipole, related to the effective magnetic moment and the dominant contribution to diffraction in most cases. The wavefunction Eq. (1) has been inferred from the energy spectrum [18]. Boothroyd has confirmed that the dipole operator alone was used to calculate transition intensities and, therefore, inter-level crystal-field transitions are not calculated correctly [34]. Even so, the wavefunction (1) is likely to be a good approximation to the true ground state. While the octupole is found to be the largest of the higher-order multipoles, it does not overwhelm. The triakontadipole, ignored in [18], is significant with $\langle \mathbf{T}^5 \rangle \sim - (1/2) \langle \mathbf{T}^3 \rangle$, and even-rank multipoles [30] should not be neglected in future discussions of the diffraction pattern.

**ACKNOWLEDGEMENTS**. SWL is grateful to Dr D. D. Khalyavin for sage advice, without which the study would not have been completed He assisted in constructing all figures apart from Fig. 2. Professor A. T. Boothroyd gave advice on the analysis of crystal-field data. Dr R.

# APPENDIX

Reduced matrix-elements used in the calculation of multipoles (3) and (4) are found in tables or calculated from formulae provided in reference [17]. We find ($Ce^{3+}$, $l = 3$),

$(lJ\|T^1\|lJ) = \sqrt{(30/7)} \{\langle j_0(\kappa)\rangle + (8/5) \langle j_2(\kappa)\rangle\}$, $(lJ\|T^1\|lJ\,') = -2\sqrt{(2/21)} \{\langle j_0(\kappa)\rangle - (1/2) \langle j_2(\kappa)\rangle\}$,

$(lJ\,'\|T^1\|lJ\,') = 8\sqrt{(2/7)} \{\langle j_0(\kappa)\rangle + (2/3) \langle j_2(\kappa)\rangle\}$,

$(lJ\|T^3\|lJ) = (12/7) \sqrt{(1/5)}\, h(\kappa)$,

$(lJ\|T^3\|lJ\,') = -(4/21) \sqrt{(2)} \{\langle j_2(\kappa)\rangle - (3/4) \langle j_4(\kappa)\rangle\}$,

$(lJ\,'\|T^3\|lJ\,') = (8/7) \sqrt{(22/3)} \{\langle j_2(\kappa)\rangle + (4/11) \langle j(\kappa)\rangle\}$,

$(lJ\|T^4\|lJ\,') = -i\,(3/2) \sqrt{(42/11)} \langle j_4(\kappa)\rangle$,

$(lJ\|T^5\|lJ) = (2/11) \sqrt{(15/7)} \{\langle j_4(\kappa)\rangle + 12 \langle j_6(\kappa)\rangle\}$,

$(lJ\|T^5\|lJ\,') = -(2/11) \sqrt{(6/7)} \{\langle j_4(\kappa)\rangle - (5/6) \langle j_6(\kappa)\rangle\}$,

$(lJ\|T^6\|lJ\,') = -i\,(5/3) \sqrt{(91/11)} \langle j_6(\kappa)\rangle$.

Here, $h(\kappa) = \{\langle j_2(\kappa)\rangle + (10/3) \langle j_4(\kappa)\rangle\}$, which is depicted in Fig. 2.

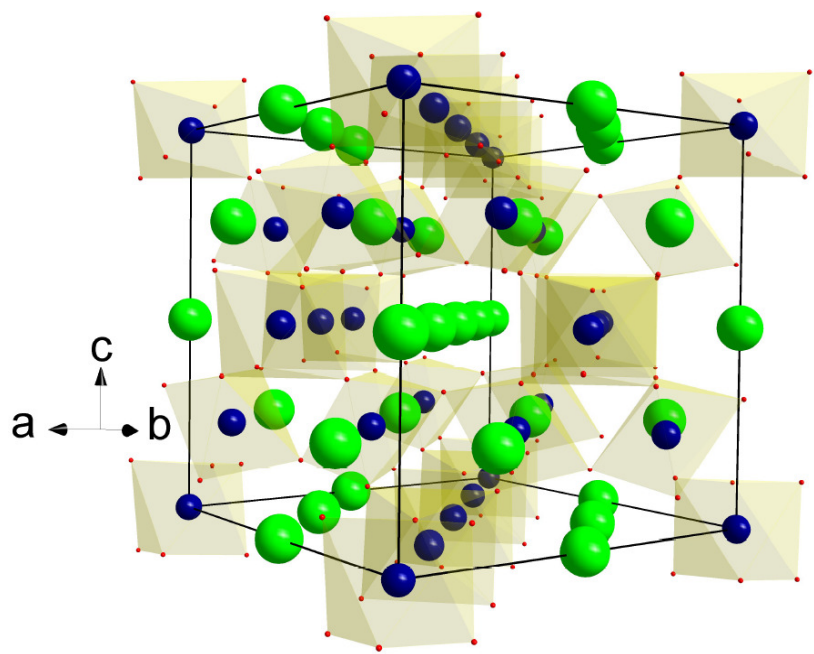

**Fig. 1.** $Ce_2Sn_2O_7$ unit cell showing 8-coordinated cerium (green), 6-coordinated tin (blue), and 4-coordinated oxygen (red).

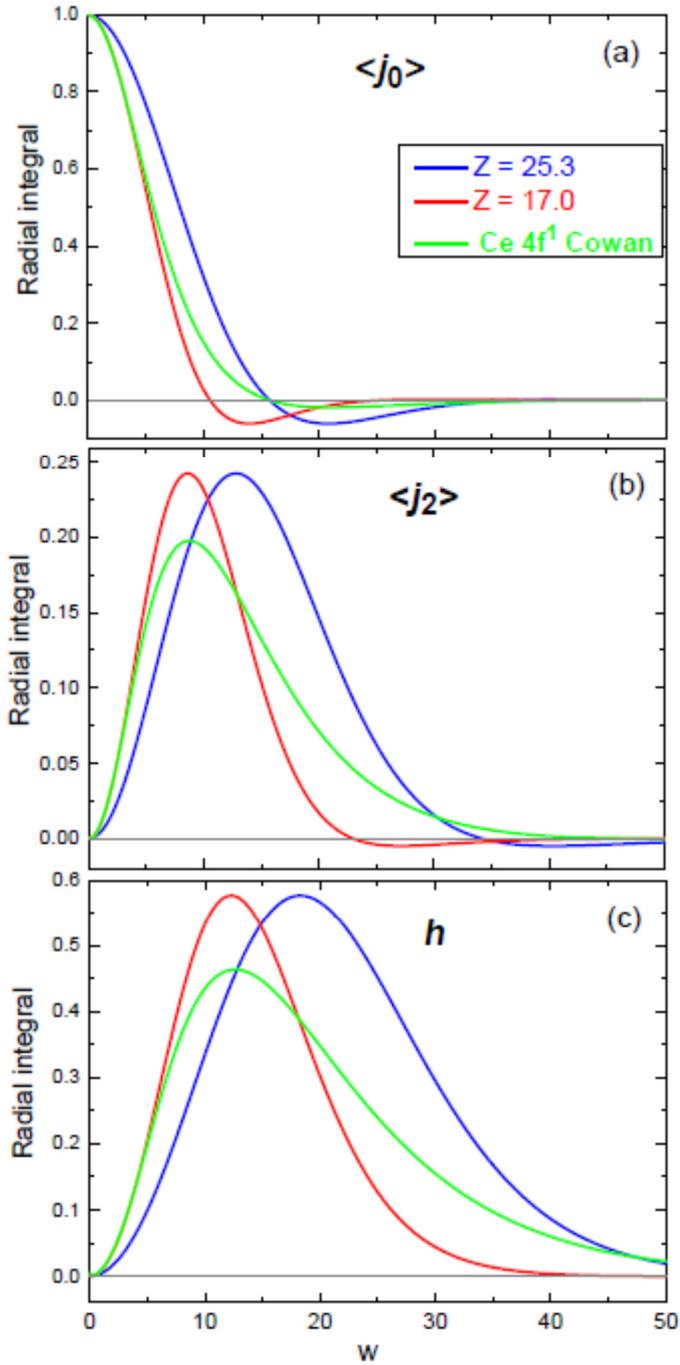

**Fig. 2**. Radial integrals for a hydrogen-like radial density Eq. (5) calculated with $Z_c = 25.3$ (blue) and $Z_c = 17.0$ (red) [18]. Corresponding quantities calculated with the atomic code are displayed in green. The radial integral $\langle j_0(\kappa) \rangle$ in (a) is unique to the dipole, which also contains $\langle j_2(\kappa) \rangle$ displayed in (b). In (c), $h(\kappa) = \{\langle j_2(\kappa) \rangle + (10/3) \langle j_4(\kappa) \rangle\}$ is the primary octupole form factor. The dimensionless parameter w and wavevector $\kappa$ are related by the Bohr radius, namely, $w = 3a_o\kappa$.

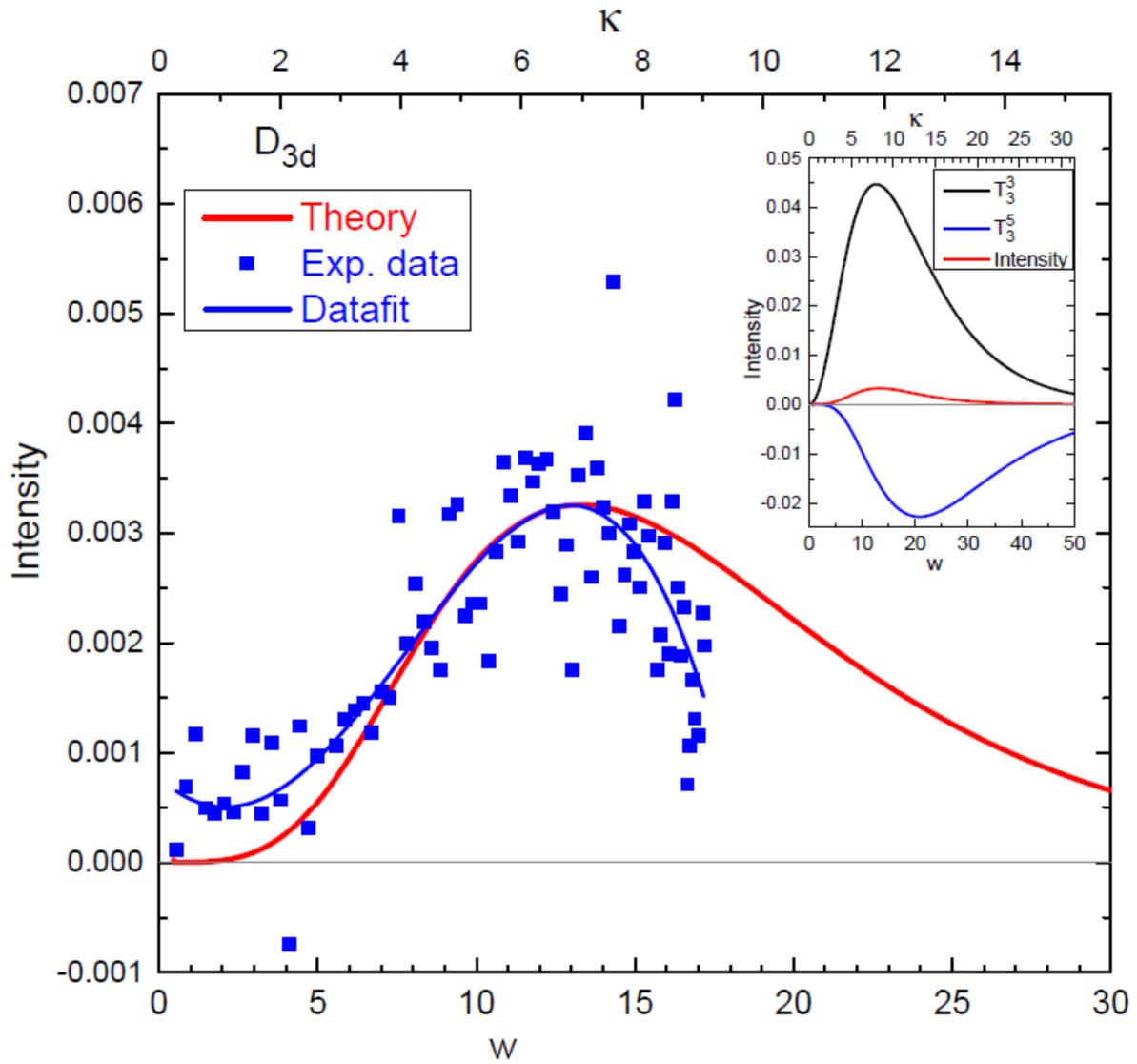

**Fig. 3**. Red curve; diffraction pattern from Eq. (6) calculated with the two multipoles in Eq. (3) defined by the point group $D_{3d}$, and our radial integrals displayed as functions of w and wavevector κ. Blue curve; polynomial fit to the data (blue squares) reported by Sibille *et al*. [18]. Imaginary part of our multipoles Eq. (3) are in the inset together with intensity.

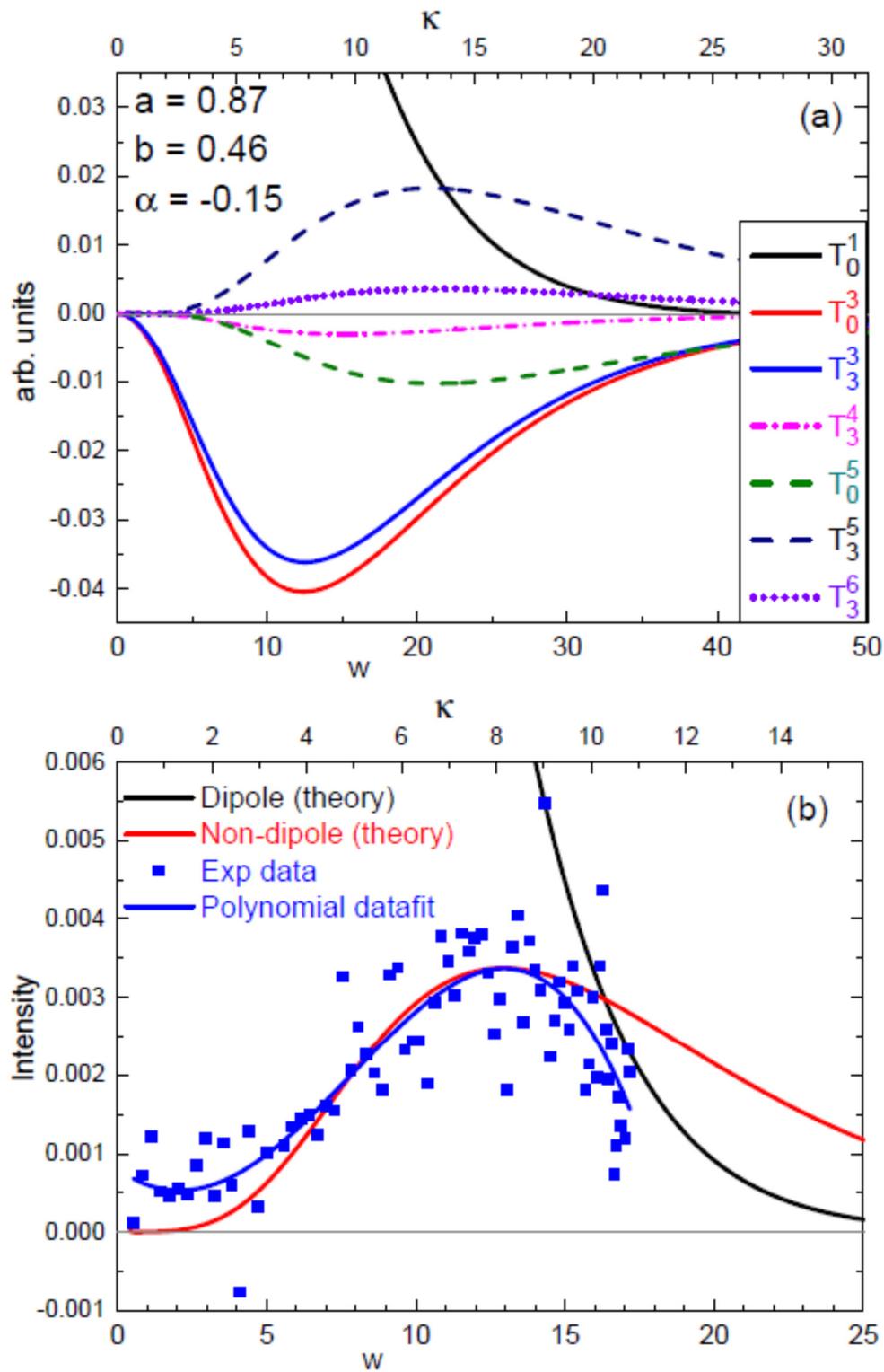

**Fig. 4**. (a) multipoles defined in Eq. (4), using the specified parameters a, b and α, that are consistent with the point group $\bar{3}m'$. (b) red curve; diffraction Eq. (6) calculated with all multipoles other than the dipole which is depicted by the black curve: blue curve; polynomial fit to the data (blue squares) reported by Sibille *et al*. [18].

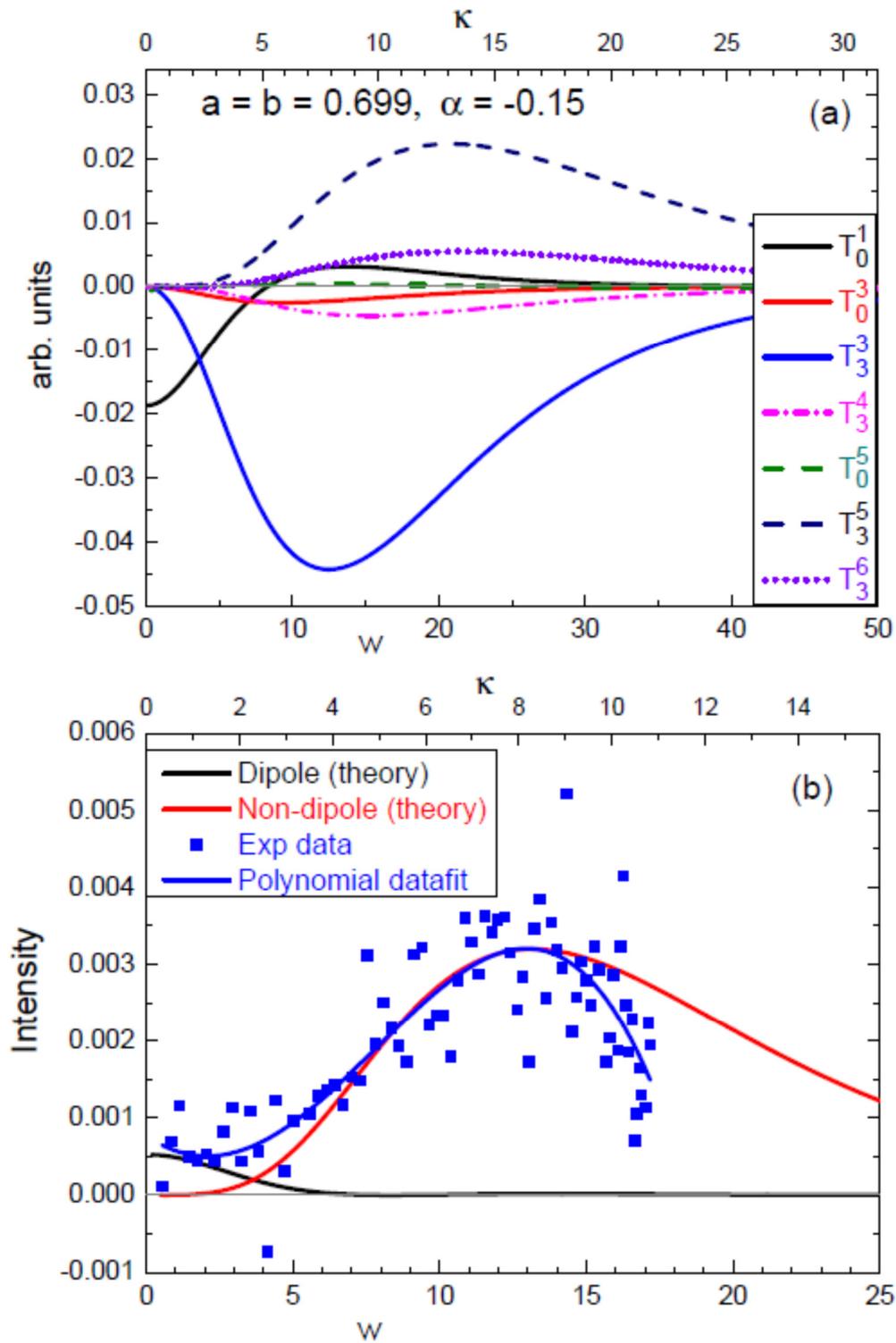

**Fig. 5**. (a) multipoles defined in Eq. (4) with a = b that defines the octupole-ice configuration [18]. (b) red curve: diffraction pattern Eq. (6) calculated with all multipoles other than the dipole: black curve; dipole contribution to intensity: blue squares and curve; experimental data and polynomial fit [18].